\RequirePackage{fix-cm}
\documentclass[smallextended]{svjour3}       
\smartqed  
\usepackage{graphicx}
\usepackage{braket}
\usepackage{multirow}

\begin{document}

\title{Efficient Controlled Quantum Secure Direct Communication Protocols
}


\author{Siddharth Patwardhan      \and    
	Subhayan Roy Moulick    \and 
        Prasanta K. Panigrahi 
}

\authorrunning{Patwardhan, Roy Moulick, Panigrahi} 

\institute{Siddharth Patwardhan \and Subhayan Roy Moulick \and Prasanta K. Panigrahi  \at
              Indian Institute of Science Education and Research Kolkata, Mohanpur 741246, West Bengal, India \\
              \email{sbp14ms080@iiserkol.ac.in, subhayan@acm.org, pprasanta@iiserkol.ac.in}           
}

\date{Received: date / Accepted: date}

\maketitle

\begin{abstract}
We study controlled quantum secure direct communication (CQSDC), a cryptographic scheme where a sender can send a secret bit-string to an intended recipient, without any secure classical channel, who can obtain the complete bit-string only with the permission of a controller. We report an efficient protocol to realize CQSDC using Cluster state and then go on to construct a (2-3)-CQSDC using Brown state, where a coalition of any two of the three controllers is required to retrieve the complete message. We argue both protocols to be unconditionally secure and analyze the efficiency of the protocols to show it to outperform the existing schemes while maintaining the same security specifications. 
\keywords{CQSDC \and (2-3)-CQSDC \and Cluster state \and Brown state}
\end{abstract}

\section{Introduction}
Establishing secure communication between two (or more) parties, without any undesirable leakage of information, is the ultimate goal of cryptography. While a conventional way of achieving this is to have a key (pair) shared among the participating parties, with at least one of the key (pair) to be private. Quantum Key Distribution (QKD) protocols \cite{GRTZ02} tend to facilitate such pursuits, in practice. Another way to accomplish this task is through Quantum Secure Direct Communication (QSDC). The first QSDC protocol, where the message is encoded in the EPR quantum states was given by Long et al., \cite{LL02}. Seminal works by Beige et al., \cite{BEKW02} and further developed by Bostr\"om et al., \cite{BF02} and Deng et al., \cite{DL02}, among others, that permits the secret messages to be transmitted directly among the desired parties without any pre-shared key. Such protocols have been well studied in the recent years, and effective protocols have been proposed utilizing single photon resources \cite{WZT06} , Bell states \cite{GYW04} \cite{DLL03} \cite{KXJZ11}, GHZ states \cite{XFLZYU06} \cite{MXA06},  GHZ-like states \cite{BP12} and W states \cite{CS06} \cite{LXYF08} \cite{LXYF08b}.

The idea of 'controlled communication', where a controller's permission is required to completely facilitate a communication, was introduced by Zutowski et al., \cite{ZZHW98}. Recents works due to \cite{ZZZ09} \cite{DXGRL11} \cite{KCH11} \cite{KH12} \cite{HH15}  \cite{P15} have motivated the study of Controlled Quantum Secure Direct Communication (CQSDC) protocols, where a message can be sent by a sender to the intended recipient, who can only recover the message, with the permission/cooperation of a controller. CQSDCs can be perceived a more general form of QSDC. Any CQSDC scheme can be reduced and made operationally equivalent to a(n) (Interactive) QSDC scheme by simply allowing the sender to assume the role of the controller in addition.
In this paper, we report a simple construction of a CQSDC scheme using Cluster state, that is feasible to implement using the current technologies. It is demonstrably shown to be more efficient than the existing protocols \cite{DXGRL11} \cite{KH12} \cite{HH15} while mainlining the same security specifications. Also unlike the existing schemes, the proposed one also does not need entanglement swapping.
We further give a construction of a (2-3)-CQSDC using Brown state, where at least two of the three controllers must form a coalition to recover the message. This is the first proposal for a thresholding based CQSDC to the best of the authors' knowledge.   

\section{CQSDC with Cluster state}
\label{sec:1}

The task of carrying out Controlled Quantum Secure Direct Communication is divided into two steps -\emph{Preparation Step} and \emph{Retrieval Step}. In the Preparation Step the sender, Alice, encodes the message using suitable local unitary transformations on her subsystem, followed by measuring her subsystems. The Retrieval Step is where the receiver, Bob, requests permission of the controller, Charlie, to recover the message. Ideally, without Charlie's permission, Bob must not be able to fully recover the message.
   
A 4-qubit Cluster state is defined as 
\begin{equation} 
\label{cluster}
\Ket{C}_{4} = \frac{ \big(\Ket{0000} + \Ket{1001}+ \Ket{0110} -\Ket{1111} \big)_{a_1,a_2,b_1,c_1}  }{2} 
\end{equation}

To send two bit messages, a four-qubit Cluster state, $\Ket{C_4}$, is used, such that two qubits are possessed by Alice, and one qubit by Bob and one qubit by Charlie, i.e., $\Ket{C}_{a_1,a_2,b_1,c_1}$ is distributed as shown in Table 1. 
\begin{center}
Table 1: Distribution of Qubits, of the Cluster state (\ref {cluster}), among the parties \\
\begin{tabular}{|c|c|c|}
\hline 
{\bf Party}& {\bf Qubits} \\ 
\hline 
Alice & $a_1, a_2$ \\ 
Bob & $b_1$ \\ 
Charlie & $c_1$ \\ 
\hline
\end{tabular}
\end{center}

In the Preparation Step, the secret message is encoded as shown in Table 2.  Here $X$ represents an operation of a $X$-Gate (Bit flip) on the qubit corresponding to the specified column, and $I$ represents an operation of a $I$-Gate (Identity) on the qubit. More concretely, $X = (\Ket{0}\Bra{1} + \Ket{1}\Bra{0})$ and $I = (\Ket{0}\Bra{0} + \Ket{1}\Bra{1})$
 
\begin{center}
Table 2: Operations Alice applies on her qubits. \\
\begin{tabular}{|c|c|c|}
\hline 
{\bf Secret Message}		& {\bf $a_1$}  & {\bf $a_2$}   \\  
\hline 
00 	&	$I$	&	$I$ \\
\hline 
01 	&	$X$	&	$X$ \\ 
\hline 
10 	&	$X$	&	$I$ \\  
\hline 
11 	&	$I$	&	$X$ \\ 
\hline
\end{tabular}
\end{center}

Following that, Alice, Bob and Charlie measure their systems and Alice sends her measurement outcomes (two c-bits) to Bob. 
During the Retrieval Step, Bob requests Charlie to send his measurement outcome to Bob. Based on the information available to Bob (i.e., Alice's measurement outcomes that Alice had sent in the Preparation Step, Bob's measurement outcomes, and the measurement outcomes received from Charlie), Bob consults the following table to recover the secret message.
\begin{center}
Table 3:  Corresponding messages for possible measurement outcomes\\
\begin{tabular}{|c|c|c|c|}
\hline
{\bf Alice's}	 & \textbf{Bob's} & \textbf{Charlie's} & \textbf{Secret }      \\ 
\textbf{Measurement} & \textbf{Measurement} & \textbf{Measurement} & \textbf{Message} \\
\hline
00              & 0             & 0                 & \multirow{4}{*}{00} \\ \cline{1-3}
01              & 1             & 0                 &                     \\ \cline{1-3}
10              & 0             & 1                 &                     \\ \cline{1-3}
11              & 1             & 1                 &                     \\ \hline \hline
00              & 1             & 1                 & \multirow{4}{*}{01} \\ \cline{1-3}
01              & 0             & 1                 &                     \\ \cline{1-3}
10              & 1             & 0                 &                     \\ \cline{1-3}
11              & 0             & 0                 &                     \\ \hline \hline
00              & 0             & 1                 & \multirow{4}{*}{10} \\ \cline{1-3}
01              & 1             & 1                 &                     \\ \cline{1-3}
10              & 0             & 0                 &                     \\ \cline{1-3}
11              & 1             & 0                 &                     \\ \hline \hline
00              & 1             & 0                 & \multirow{4}{*}{11} \\ \cline{1-3}
01              & 0             & 0                 &                     \\ \cline{1-3}
10              & 1             & 1                 &                     \\ \cline{1-3}
11              & 0             & 1                 &                     \\ \hline
\end{tabular}
\end{center}

To exemplify, suppose Alice wants to send a message $01$ to Bob, then following the distribution of qubits as in Table 1, Alice applies an $X\circ X$ to her system (as in Table 2) and measures it and reports the measurement outcome, say $b_1, b_2$, using an authenticated channel, to Bob. Bob now requests Charlie for his \emph{permission}. Charlie measures his system forwards his measurement outcome, say $b_4$ to Bob. Bob also measures his system to see $b_3$. Bob now looks up Table 3 to identify the column for the tuple $(b_1,b_2,b_3,b_4)$ and be able to recover the message $01$.

\subsection{Security Analysis}
	The security analysis of the protocol includes the analysis of the scheme without the permission of the controller and the condition, where all the classical information is displayed to Eve.

\subsubsection{Analysis of the scheme without the permission of Charlie}
Notice the distribution of classical bits due to measurement outcomes over the space messages (as in Table 3) is uniform in the length of message. Thus, in this case, Bob must need all $4$ bits of classical information to identify the intended message sent to him. Without the permission of Charlie, he only has access to atmost $3$ bits of information. If Bob has a better strategy than to guess Charlie's bit, with a guessing probability $> 1/2$, then this violated Information Causality \cite{PPKS09}, which says the maximum information that Bob can gain about previously unknown knowledge, using all local resources and n classical bits, is at most n. However, with access to just Alice's measurement outcomes, and his own measurement outcome, Bob does learn about learn 1 bit of information. It can be seen, this $1-bit$ information that Bob learns is the 2nd bit of the message.
   
Thus, the probability that Bob can correctly guess the information Alice has sent without the permission of Charlie is given by,
$$Pr(mn=00|xyz)=Pr(mn=01|xyz)=Pr(mn=10|xyz)=Pr(mn=11|xyz)=1/2$$	
where $mn$, $xy$ and $z$ indicate the secret message, Alice's reading and Bob's reading respectively.

\subsubsection{Analysis of the scheme when all the classical information is displayed}
The same argument as in Section 2.1.1. can be used to convince the inability of an Evesdropper, Eve, who has access to all classical communication from Alice and Charlie, to predict the classical message. Due to the uniformity in the view of Bob and Charlie, the roles of the Eve and Bob in here can be perceived as Bob and Charlie in 2.1.1. The 1 bit of information that Eve learns here is the parity of the message. 

Thus, the probability that Eve can correctly guess the information when she has access to all the classical bits sent is given by,
$$Pr(mn=00|xyz)=Pr(mn=01|xyz)=Pr(mn=10|xyz)=Pr(mn=11|xyz)=1/2$$
where $mn$, $xy$ and $z$ indicate the secret message, Alice's reading and Charlie's reading respectively.

\subsection{Efficiency}
The efficiency of the protocols can be measured using the following equations, as previously used in \cite{DXGRL11}, \cite{KCH11},  \cite{HH15}.
\begin{equation}
\label{eff1}
\eta_{1} = \frac{m_{u}}{q_{k}+b{k}}
\end{equation}
\begin{equation}
\label{eff2}
\eta_{2} = \frac{m_{u}}{q_{k}}
\end{equation}

where $m_{u}$, $q_{k}$ and $c_{k}$ denote the number of bits in the secret message, number of qubits used and the number of bits of classical communication used. Ideally, higher the value of $\eta_{1}$ and $\eta_{2}$, the more efficient the scheme would be.

The values of the $\eta_{1}$ and $\eta_{2}$ for the proposed protocol along with the other protocols are given in the Table 4.

\begin{center}
Table 4: Comparison of efficiency of various protocols \\  
\begin{tabular}{|c|c|c|}
\hline
{\bf Protocol} & {\bf $\eta_{1}$} & {\bf $\eta_{2}$} \\
\hline
Dong et al. \cite{DXGRL11} & 0.125 & 0.25 \\
Kao et al.  \cite{KCH11} & 0.125 & 0.25 \\
Hassanpour et al. \cite{HH15} & 0.22 & 0.33 \\
Proposed Protocol & 0.28 & 0.50 \\
\hline
\end{tabular}
\end{center}
It may be noted, unlike the other protocols, the proposed protocol doesn't make use of Entanglement Swapping or other complex operations, however, as demonstrated, attains better efficiency than the existing ones. Also compared to other protocols, the proposed protocol is simpler to implement and is inexpensive in terms of the number of gates and measurements required to realize this.

\section{(2-3)-CQSDC with Brown state}
\label{sec:1}

A (2-3)-CQSDC is a thresholding based Controlled Quantum Secure Direct Communication, where a sender sends a two-bit message, that can be completely recovered only when any two of three controllers form a coalition.  To realize the protocol, a 5-qubit Brown state is used here. 

\begin{equation} 
\label{brown}
\Ket{B}_5 = \frac{ \big( \Ket{001}\Ket{\Phi^{-}}+ \Ket{010}\Ket{\Psi^{-}}+ \Ket{100}\Ket{\Phi^{+}}-\Ket{111}\Ket{\Psi^{+} } \big)_{a,b,c,d,e} }{2}
\end{equation}

where $\Ket{\Phi^{\pm}}= \frac{1}{\sqrt{2}} ( \Ket{00} \pm \Ket {11} )$ and $\Ket{\Psi^{\pm}}= \frac{1}{\sqrt{2}} (\Ket{01} \pm \Ket {10})$. Brown state \cite{BSSB05} have shown their utility to carry out diverse quantum tasks, including teleportation, quantum state sharing, superdense coding have been studied rather exhaustively \cite{MP08}. Here, to perform a (2-3) CQSDC, a 5-qubit Brown state, $\Ket{B}_{a,b,c,d,e}$ is distributed among the concerned parties - the sender, Alice, and the controllers, Charlie1, Charlie2, Charlie3, as in Table 5. 
\begin{center}
Table 5: Distribution of Qubits of the Brown state (\ref{brown}) among the parties \\
\begin{tabular}{|c|c|}
\hline 
{\bf Party}& {\bf Qubits} \\ 
\hline 
Alice & $d,e$ \\ 
Charlie1 & $a$ \\ 
Charlie2 & $b$ \\ 
Charlie3 & $c$ \\ 
\hline
\end{tabular}
\end{center}

In the Preparation Step, to encode a two bit message, Alice applies the suitable unitaries on her subsystem as described in the Table 6, where $X$ represents an operation of a $X$-Gate (Bit flip) on the qubit, $I$ represents an operation of a $I$-Gate (Identity) on the qubit and $Z$ represents an operation of a $Z$-Gate(phase flip) on the qubit. More concretely,  $I = (\Ket{0}\Bra{0} + \Ket{1}\Bra{1})$, $X = (\Ket{0}\Bra{1} + \Ket{1}\Bra{0})$,  $Z = (\Ket{0}\Bra{0} - \Ket{1}\Bra{1})$

\begin{center}
Table 6: Operations Alice applies on her qubits. \\
\begin{tabular}{|c|c|c|}
\hline 
{\bf Secret Message}		& {\bf $a_1$}  & {\bf $a_1$}   \\  
\hline 
00 	&	$I$	&	$I$  \\
\hline 
01 	&	$Z$	&	$X$ \\ 
\hline 
10 	&	$Z$	&	$I$  \\  
\hline 
11 	&	$I$	&	$X$  \\ 
\hline
\end{tabular}
\end{center}

Following that, Alice, and the controllers, Charlie1, Charlie2, and Charlie3 measure their qubits, and Alice sends her measurement outcomes to all the controllers.

In the Retrieval Step, at least two of the three controllers, must come together and collaborate to get the secret message. Based on the distribution as in Table 7, it may be noted, given Alice's Measurements are known to all, the coalition of two controllers is necessary and sufficient to recover the secret message.

It may be noted that it is possible  for any of the controllers to presume the role of Bob during the protocol. It is also possible for at-most two of the three controllers to recover the message, privately with the permission of another controller.  Suppose Alice wants to send a message $10$. Following the distribution of qubits as in Table 5, Alice applies the operation $Z \otimes X$ to her subsystem (as in Table 6) and measures it to see an outcome, say$b_1,b_2$ and sends her outcomes to the intended recipient (say Charlie$_i$). Charlie$_i$ now measures his system, to see $b_3$ and with the \emph{permission} of Charlie$_j, (i\neq j)$, who, on consent, measures his own system to see and correctly reports $b_4$, now has a tuple $(b_1, b_2, b_3, b_4)$ can recover the intended message Alice had send, by looking up Table 7.
\vspace{10em}
\begin{center}
Table 7:  Corresponding messages for possible measurement outcomes\\
\begin{tabular}{|c|c|c|c|c|}
\hline
\scriptsize{{\bf Alice's}}	 		& \scriptsize{\textbf{Charlie1's}} 			& \scriptsize{\textbf{Charlie2's}} 		& \scriptsize{\textbf{Charlie3's}} 	& \scriptsize{\textbf{Secret }}     \\ 
\scriptsize{\textbf{Measurement}} 	& \scriptsize{\textbf{Measurement}} 		& \scriptsize{\textbf{Measurement}} 	& \scriptsize{\textbf{Measurement}}	& \scriptsize{\textbf{Message}} \\
\hline
$\phi^-$ 	& 0             & 0		& 1         	& \multirow{4}{*}{00} \\ \cline{1-4}
$\psi^-$ 	& 0             & 1		& 0 		&                     \\ \cline{1-4}
$\phi^+$ 	& 1             & 0		& 0		&                     \\ \cline{1-4}
$\psi^+$	& 1             & 1            & 1	     	&                     \\ \hline \hline

$\psi^-$ 	& 0             & 0		& 1         	& \multirow{4}{*}{01} \\ \cline{1-4}
$\phi^-$ 	& 0             & 1		& 0 		&                     \\ \cline{1-4}
$\psi^+$ 	& 1             & 0		& 0		&                     \\ \cline{1-4}
$\phi^+$	& 1             & 1            & 1	     	&                     \\ \hline \hline

$\phi^+$ 	& 0             & 0		& 1         	& \multirow{4}{*}{10} \\ \cline{1-4}
$\psi^+$	& 0             & 1		& 0 		&                     \\ \cline{1-4}
$\phi^-$ 	& 1             & 0		& 0		&                     \\ \cline{1-4}
$\psi^-$	& 1             & 1            & 1	     	&                     \\ \hline \hline

$\psi^+$ 	& 0             & 0		& 1         	& \multirow{4}{*}{11} \\ \cline{1-4}
$\phi^+$ 	& 0             & 1		& 0 		&                     \\ \cline{1-4}
$\psi^-$ 	& 1             & 0		& 0	&                     \\ \cline{1-4}
$\phi^-$	& 1             & 1            & 1	     	&                     \\ \hline
\end{tabular}
\end{center}

\subsection{Security Analysis}
\subsubsection{Analysis of the scheme without the permission of a second controller}
Notice the distribution of classical bits due to measurement outcomes of Alice and any two Charlie$_i$ and Charlie$_j$, over space space of messages is uniform, as in Table 7. Hence to for a Charlie$_k$, recover the message, he needs 4 bits of information. Without the permission of a second controller, Charlie$_k$ has access to only 3 bits of information. If Bob could design a better strategy than randomly guessing the second controller's measurement outcome with probability greater than $\frac{1}{2}$, then he violates Information Causality \cite{PPKS09}, in the same spirit as argued in Section 2.1.1.
Hence,  The probability that only one of the three can correctly guess the information Alice has sent without the permission of a second controller is given by,
 $$Pr(mn=00|xyz)=Pr(mn=01|xyz)=Pr(mn=10|xyz)=Pr(mn=11|xyz)=1/2$$	
where $mn$ is the secret message, $xy$ is Alice's measurement outcome and $z$ is the measurement outcome of Charlie$_k$.

\subsubsection{Analysis of the scheme when all the classical information is displayed}
An eavesdropper, Eve, who sees the classical communication between Alice and one of the Charlie's, say Charlie$_i$, has access to 3 bits of information. Of-course, if both two of the controller's, Charlie$_j$, Charlier$_k$ in the protocol want to help Charlie$_i$, Eve learns 4 bits of information, and can figure out the intended message. However this can be avoided, by letting only one controller to control the permission. In a case where both controllers want to help, only one must send out the classical bits. Now, given 3 bits of information to Eve, one can build up the same argument as in Section 3.1.1, with Eve and Charlie$_i$ (here) playing the role of Charlie$_i$ and Charlie$_{j or k}$ (as in 3.1.1), to claim, given 3-bits of information an eavesdropper can figure out the message, then it violates Information Causality \cite{PPKS09}.

Hence, The probability that Eve can correctly guess the information when she has access to all the classical bits sent is given by,
$$Pr(mn=00|xyz)=Pr(mn=01|xyz)=Pr(mn=10|xyz)=Pr(mn=11|xyz)=1/2$$
where $mn$, $xy$ and $z$ indicate the secret message, Alice's reading and Charlie's reading respectively.

\subsection{Efficiency} 
The values of $\eta_{1}$ and $\eta_{2}$ for the proposed protocol, computed based on Eq. \ref{eff1} and Eq. \ref{eff2} are 0.25 and 0.40.

The exquisite nature of the five qubit Brown state, where two of the particles are entangled as Bell States makes it possible to use Brown state in such a conditionally controlled communication scheme.      

\section{Conclusion}
We have presented here two efficient protocols for Controlled Quantum Secure Direct Communication. The first one is an unidirectional CQSDC scheme, involving a sender, controller and a receiver, that makes use of Cluster state, where the sender sends two bit messages to the receiver who recovers the message with the help of a controller. We analyze the security and efficiency of the proposed scheme and report it to be more efficient than the existing protocols while maintaining the same security specifications. 
Following that, we introduce a thresholding based (2-3)-CQSDC scheme using Brown state, involving a sender and three controllers, where only two controllers need to collude to recover a two bit message sent by a sender.


\bibliographystyle{spphys}       
\bibliography{references}   

\begin{thebibliography}{10}
\providecommand{\url}[1]{{#1}}
\providecommand{\urlprefix}{URL }
\expandafter\ifx\csname urlstyle\endcsname\relax
  \providecommand{\doi}[1]{DOI \discretionary{}{}{}#1}\else
  \providecommand{\doi}{DOI \discretionary{}{}{}\begingroup
  \urlstyle{rm}\Url}\fi

\bibitem{GRTZ02}
N.~Gisin, G.~Ribordy, W.~Tittel, H.~Zbinden, Reviews of modern physics
  \textbf{74}(1), 145 (2002)

\bibitem{LL02}
G.L. Long, X.S. Liu, Physical Review A \textbf{65}(3), 032302 (2002)

\bibitem{BEKW02}
A.~Beige, B.G. Englert, C.~Kurtsiefer, H.~Weinfurter, Acta Phys. Pol. A
  \textbf{101}(357) (2002)

\bibitem{BF02}
K.~Bostr{\"o}m, T.~Felbinger, Physical Review Letters \textbf{89}(18), 187902
  (2002)

\bibitem{DL02}
F.G. Deng, G.L. Long, Physical Review A \textbf{69}(5), 052319 (2004)

\bibitem{WZT06}
J.~Wang, Q.~Zhang, C.j. Tang, Physics Letters A \textbf{358}(4), 256 (2006)

\bibitem{GYW04}
T.~Gao, F.L. Yan, Z.X. Wang, Il Nuovo Cimento \textbf{119B}(313) (2004)

\bibitem{DLL03}
F.G. Deng, G.L. Long, X.S. Liu, Physical Review A \textbf{68}(4), 042317 (2003)

\bibitem{KXJZ11}
L.~Kai, H.~Xiao-ying, T.~Ji-hong, L.~Zhen-hua, in \emph{Multimedia Information
  Networking and Security (MINES), 2011 Third International Conference on}
  (IEEE, 2011), pp. 73--76

\bibitem{XFLZYU06}
Y.~Xia, C.B. Fu, F.Y. Li, S.~Zhang, K.H. Yeon, C.I. Um, Journal of Korean
  Physical Society \textbf{47}

\bibitem{MXA06}
Z.X. Man, Y.J. Xia, N.B. An, Journal of Physics B: Atomic, Molecular and
  Optical Physics \textbf{39}(18), 3855 (2006)

\bibitem{BP12}
A.~Banerjee, A.~Pathak, Physics Letters A \textbf{376}(45), 2944 (2012)

\bibitem{CS06}
H.J. Cao, H.S. Song, Physica Scripta \textbf{74}(5), 572 (2006)

\bibitem{LXYF08}
D.~Li, X.~Xiao-Ming, G.~Ya-Jun, C.~Feng, Communications in Theoretical Physics
  \textbf{49}(6), 1495 (2008)

\bibitem{LXYF08b}
D.~Li, X.~Xiao-Ming, G.~Ya-Jun, C.~Feng, Communications in Theoretical Physics
  \textbf{50}(2), 359 (2008)

\bibitem{ZZHW98}
M.~{\.Z}ukowski, A.~Zeilinger, M.~Horne, H.~Weinfurter, Acta Physica Polonica A
  \textbf{93}(1), 187 (1998)

\bibitem{ZZZ09}
L.L. Zhang, Y.B. Zhan, Q.Y. Zhang, International Journal of Theoretical Physics
  \textbf{48}(10), 2971 (2009)

\bibitem{DXGRL11}
L.~Dong, X.M. Xiu, Y.J. Gao, Y.P. Ren, H.W. Liu, Optics Communications
  \textbf{284}(3), 905 (2011)

\bibitem{KCH11}
S.H. Kao, T.~Chia-Wei, T.~Hwang, Communications in Theoretical Physics
  \textbf{55}(6), 1007 (2011)

\bibitem{KH12}
S.H. Kao, T.~Hwang, Proc. of International Conference on Information Security
  and Intelligence Control pp. 141--145 (2012)

\bibitem{HH15}
S.~Hassanpour, M.~Houshmand, Quantum Information Processing \textbf{14}(2), 739
  (2015)

\bibitem{P15}
A.~Pathak, Quantum Information Processing \textbf{14}(6), 2195 (2015)

\bibitem{PPKS09}
M.~Paw{\l}owski, T.~Paterek, D.~Kaszlikowski, V.~Scarani, A.~Winter,
  M.~{\.Z}ukowski, Nature \textbf{461}(7267), 1101 (2009)

\bibitem{BSSB05}
I.D. Brown, S.~Stepney, A.~Sudbery, S.L. Braunstein, Journal of Physics A:
  Mathematical and General \textbf{38}(5), 1119 (2005)

\bibitem{MP08}
S.~Muralidharan, P.K. Panigrahi, Physical Review A \textbf{77}(3), 032321
  (2008)

\end{thebibliography}


\end{document}